\documentclass[prb,twocolumn,showpacs]{revtex4}

\usepackage{graphicx}
\usepackage{amsmath}
\usepackage{amsfonts}
\usepackage{bm}
\usepackage{bbm}

\begin{document}

%......Title and other stuff
\title{Spin-orbit coupling and crystal-field splitting in the electronic
and optical properties of nitride quantum dots with a wurtzite
crystal structure}

%\author{S. Schulz\inst{1}\thanks{\email{sschulz@itp.uni-bremen.de}} \and S.
%Schumacher\inst{2} \and G. Czycholl\inst{1}}

%\institute{Institute for Theoretical Physics, University of Bremen,
%28359 Bremen, Germany \and College of Optical Sciences, University
%of Arizona, Tucson, Arizona 85721, USA}

%\titlerunning{Spin-orbit coupling and crystal-field splitting in the electronic
%and optical properties of nitride QDs}
\author{S. Schulz}
\affiliation{Institute for Theoretical Physics,
             University of Bremen,
             28359 Bremen, Germany}
\author{S. Schumacher}
\affiliation{College of Optical Sciences,
             University of Arizona, Tucson,
             Arizona 85721, USA}

\author{G. Czycholl}
\affiliation{Institute for Theoretical Physics,
             University of Bremen,
             28359 Bremen, Germany}

\date{\today}


\begin{abstract}
We present an $sp^3$ tight-binding model for the calculation of the
electronic and optical properties of wurtzite semiconductor quantum
dots (QDs). The tight-binding model takes into account strain,
piezoelectricity, spin-orbit coupling and crystal-field splitting.
Excitonic absorption spectra are calculated using the configuration
interaction scheme. We study the electronic and optical properties
of InN/GaN QDs and their dependence on structural properties,
crystal-field splitting, and spin-orbit coupling.
\end{abstract}

%\PACS{78.67.Hc \and 73.22.Dj \and 71.35.-y}

\maketitle

\section{Introduction}

Group-III nitrides have shown great potential in optoelectronic
devices with a wide range of applications.~\cite{Ambach98}
\mbox{InGaN/GaN} quantum wells typically constitute the active
region in light-emitting diodes and laser diodes, covering a wide
spectral range from near ultraviolet to infrared. In particular,
devices with an active region based on pure or almost pure InN are
of great interest to reach operational frequencies in the infrared
spectral range.~\cite{GhHa2003} The small InN band gap (0.7-0.8
eV)~\cite{VuMe2003} can be extremely useful for
telecommunication-wavelength devices. The combination of this
inherent property of InN with the self-assembly of nanostructures
based on this material, provides further possibilities for useful
future optoelectronic devices. In particular, quantum dot (QD)
structures are promising candidates, acting as electron-hole
recombination centers increasing the emission efficiency. These
zero-dimensional nanostructures also have the potential to act as
single-photon emitting devices.~\cite{MiKi2000}

This work is dedicated to the investigation of the electronic and
optical properties of self-assembled InN QDs. The present study is
based on a fully atomistic empirical tight-binding model. In
contrast to multi-band $\mathbf{k}\cdot\mathbf{p}$ approaches it
takes into account the structure of the underlying atomic lattice
and is capable to describe the electronic wave functions beyond an
envelope function approximation. Our model includes strain effects
and electrostatic built-in fields, which are of major importance in
group-III nitride based nanostructures with an underlying wurtzite
crystal structure. In this paper, special attention is paid to the
possible influence of the weak crystal-field (CF) splitting and weak
spin-orbit (SO) coupling on the localized single-particle wave
functions as well as on the optical properties of the investigated
structures. These effects have commonly been neglected in previous
studies, some of which were dedicated to GaN/AlN
nanostructures~\cite{AnORe2000,AnORe2001,FoBa2003,SaAr2003}, others
to InN/GaN
nanostrucutres\cite{SaAr2002a,SaAr2002,BaSc2005,ScSc2006,BaSc2007,ScSc2007}.
Since in these materials both effects are merely of the order of a
few meV (Ref.~\cite{VuMe2003}), this approximation can be well
justified.

On a more fundamental level the resulting symmetry properties
dictated by spatial and spin degrees of freedom and, taking SO
coupling into account, by the coupled influence of both, have the
potential to support or lift certain degeneracies in the electronic
energy spectra. Based on group-theoretical arguments we discuss that
in the system under investigation for both electrons and holes, at
most two-fold degeneracies are supported. This is especially of
interest with regard to a recent discussion of the energy level
structure in semiconductor QDs with a zinc blende
structure~\cite{BeZu2005} and recent results for wurtzite InGaN/GaN
QDs obtained within an 8-band $\mathbf{k}\cdot\mathbf{p}$
model\cite{WiSc2006}. Following these general group-theoretical
arguments, our numerical results show that SO coupling and CF
splitting slightly change the results for the InN/GaN QD system
under investigation. However, the additional splittings in the
electronic energy shell structure caused by the SO coupling are at
most of the order of a few meV. From this, no significant
qualitative changes of the optical properties are found: The
excitonic absorption lines show only very small additional
splittings and the interband dipole selection rules are basically
unaltered.

\section{Theory and quantum dot model}

\subsection{The Tight-Binding Model}
\label{sec:CFSO}

For the investigation of the single particle states in small QD
structures, the description by a multi-band approach is required. In
order to take into account the underlying wurtzite structure of the
structures under investigation, we choose a microscopic $sp^3$
tight-binding (TB) model. The general aspects of the TB model are
discussed in detail in Ref.~\cite{ScSc2006}. Here, we briefly
summarize the main ingredients of this model, and focus our
attention on the role of crystal-field splitting and spin-orbit
coupling.

In the $sp^3$ TB model of Ref.~\cite{ScSc2006} the relevant
electronic structure of anions and cations is, for each spin
orientation, described by the outermost valence orbitals, $s$,
$p_x$, $p_y$ and $p_z$, and the overlap of these basis orbitals is
restricted to nearest neighbors. Being only of the order of a few
meV, so far, the influence of CF splitting and SO coupling has been
neglected in the model. In the present work, we extend our TB
model~\cite{ScSc2006} to introduce these two contributions and
investigate their possible influence on the electronic
single-particle states and energies in InN/GaN QD systems.

\begin{figure}[t]
\centering
\includegraphics{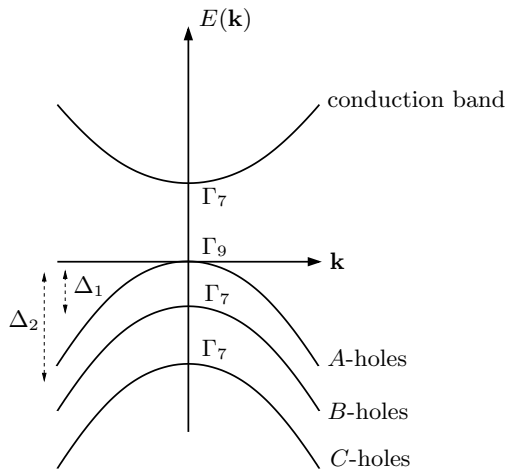}
\caption{Schematic band structure of wurtzite semiconductors with
conduction band and three valence bands. The valence band splittings
introduced by crystal field splitting and spin-orbit coupling are
denoted by $\Delta_1$ and $\Delta_2$, respectively. $\Delta_1$ and
$\Delta_2$ are calculated according to Eq.~(\ref{Eq:SplitNoStrain}).
Additionally, the symmetries of the irreducible representations
$\Gamma_i$ are given.} \label{fig:schematicBandstruct}
\end{figure}

The combination of CF splitting and SO interaction leads to a
so-called three-edge structure in the vicinity of the $\Gamma$
point. The top of the resulting valence band structure is commonly
labeled as $A$, $B$, and $C$ bands in order of increasing energy.
This three-edge structure is schematically shown in
Fig.~\ref{fig:schematicBandstruct}. Two of these three bands are of
$\Gamma_7$ and one of $\Gamma_9$ symmetry. To describe this band
structure in the vicinity of the $\Gamma$ point in the framework of
a TB model one has to take into account both CF splitting and SO
interaction.

\begin{table}
\centering \caption[Tight-binding parameters for wurtzite InN and
GaN.]{\label{TBparametersInNwurtz}Tight-binding parameters (in eV)
for the nearest neighbors of wurtzite InN and GaN. The notation of
Ref.~\cite{KoSa83} is used.}
\begin{tabular}{|l|c|c|c||c|c|c|}
\hline\hline  & \multicolumn{3}{|c||}{InN [eV]} & \multicolumn{3}{c|}{GaN [eV]} \\
\hline
& $\Delta_{\text{cf}}=0$ & $\Delta_{\text{cf}}\neq0$ & $\Delta_{\text{cf}}\neq0$ & $\Delta_{\text{cf}}=0$ & $\Delta_{\text{cf}}\neq0$ & $\Delta_{\text{cf}}\neq0$\\
& $\Delta_{\text{so}}=0$ & $\Delta_{\text{so}}=0$ & $\Delta_{\text{so}}\neq0$ & $\Delta_{\text{so}}=0$ & $\Delta_{\text{so}}=0$ & $\Delta_{\text{so}}\neq0$\\
\hline
E(s,a) & -6.791 & -6.5134 & -6.6046 & -11.012 & -8.9893 & -8.5282\\
E(p,a) & 0.000 & 0.0000 & 0.0000 & 0.005 & 0.0015 & -0.0024\\
E($\text{p}_z$,\text{a}) & 0.000 & -0.0418 & -0.0400 & 0.005 & -0.0203 & -0.0208\\
E(s,c) & -3.015 & -3.3923 & -3.3500 & 1.438 & 0.7851 & 0.6945\\
E(p,c) & 8.822 & 8.8220 & 8.8203 & 10.896 & 10.0986 & 10.0996\\
V(s,s) & -5.371 & -5.5267 & -5.5330 & -5.318 & -5.6918 & -5.6808\\
V(x,x) & 0.022 & 0.0156 & 0.1221 & -0.222 & -0.1223 & -0.0699\\
V(x,y) & 6.373 & 6.3794 & 6.2772 & 7.136 & 6.7902 & 6.7328\\
V(sa,pc) & 0.370 & 0.9576 & 0.9307 & 0.628 & 0.2641 & 1.3633\\
V(pa,sc) & 7.5 & 7.5574 & 7.4136 & 7.279 & 8.0324 & 7.7173\\
$\lambda$ & 0 & 0 & 0.0016 & 0 & 0 & 0.0023\\
\hline\hline
\end{tabular}
\end{table}

\begin{figure}[b]
\centering
\includegraphics{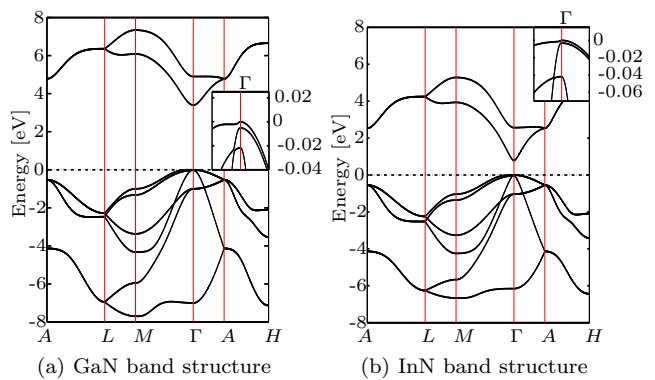}
\caption[Wurtzite bulk band structure of InN  and GaN including
crystal field splitting and spin-orbit coupling.]{(Color online)
Bulk band structure of wurtzite (a) InN  and (b) GaN obtained using
an $sp^3$ TB model including crystal field splitting and spin-orbit
coupling. The insets show the three-edge valence band structure in
the vicinity of the $\Gamma$ point.} \label{fig:bulkbandCFSO}
\end{figure}

The SO coupling is included as outlined by Chadi in
Ref~\cite{Chadi77}. Due to the high ionicity of the bonds in the
nitride system~\cite{YoXu2002}, the contribution of the SO coupling
to the valence band structure is dominated by the anion
contributions. Therefore, by introducing the parameter $\lambda$, we
include spin-orbit coupling at the anion sites only. The parameter
$\lambda$ is used to reproduce the correct splitting $\Delta_1$ of
the valence bands $A$ and $B$.

As discussed in Ref.~\cite{KoSa83}, the small CF splitting
$\Delta_{\text{cf}}$ of the wurtzite crystal differentiates the
$p_z$ orbital from the $p_x$ and $p_y$ orbitals. Pseudo-potential
calculations in local density approximation indicate that for the
studied materials the bulk crystal field splitting between the $A$
and $C$ valence bands, schematically shown in
Fig.~\ref{fig:bulkbandCFSO}, cannot be reproduced from first
principles, unless third-nearest-neighbor interactions are taken
into account~\cite{MuNa94}. The TB model discussed in
Ref.~\cite{ScSc2006} considers only nearest-neighbor hopping matrix
elements and treats the four nearest neighbor atoms as equivalent.
To account for the $A-C$ splitting within the empirical $sp^3$ TB
model with nearest-neighbor coupling, we introduce the additional
parameter $E^{A}_{p_z,p_z}$ on the anion sites for the on-site
matrix elements of the $p_z$ orbitals. This additional term is used
to reproduce the splitting $\Delta_2$ of the $A-C$ bands at the zone
center $\Gamma$.

With four atoms per unit cell, the resulting Hamiltonian is a
$32\times32$ matrix for each $\mathbf{k}$-point. This Hamiltonian
parametrically depends on the different TB matrix elements. The
parameters given in Ref.~\cite{ScSc2006} have been re-calculated to
reproduce the three-edge structure the wurtzite band
structure~\cite{FrSc2004,ZhBa99,VuMe2003} in the vicinity of the
$\Gamma$ point. The additional matrix elements $\lambda$ and
$E^{A}_{p_z,p_z}$ are adjusted to reproduce the splittings between
the different valence bands ($A$, $B$ and $C$ bands), which are
given by~\cite{Grundmann2006}
\begin{equation}
\Delta_{1,2}=\left(\frac{\Delta_\text{so}+\Delta_\text{cf}}{2}\right)\mp\sqrt{\left(\frac{\Delta_\text{so}+\Delta_\text{cf}}{2}\right)^2-\frac{2}{3}\Delta_{\text{so}}\Delta_{\text{cf}}}\,\,
, \label{Eq:SplitNoStrain}
\end{equation}
where SO and CF splitting are denoted as $\Delta_\text{so}$ and
$\Delta_\text{cf}$ respectively. Table~\ref{TBparametersInNwurtz}
summarizes the resulting TB parameters. The bulk band structures
obtained from these parameters are shown in
Fig.~\ref{fig:bulkbandCFSO} for GaN and InN. The `complicated'
valence band structure in the vicinity of the $\Gamma$ point shows
very good agreement with other TB models~\cite{JaBa2002} (cf. insets
of Fig.~\ref{fig:bulkbandCFSO}), ab-initio
approaches~\cite{CaWe2005}, pseudo potential and
$\mathbf{k}\cdot\mathbf{p}$ calculations~\cite{PuDu99}.

\subsection{Geometry of the Quantum Dot Structure} \label{sec:geometryInN}

In the framework of a TB model, the QD is modeled on an atomistic
level. The TB parameters at each atom site $\mathbf{R}$ of the
underlying wurtzite lattice are set according to the bulk values of
the respective occupying atom. In the following, we consider
lens-shaped InN QDs, grown in $c$ direction and residing on an InN
wetting layer (WL). The entire structure is embedded inside a GaN
matrix. For the numerical calculations, a finite cell (box) with a
wurtzite lattice and fixed (zero) boundary conditions is used. The
cell is sufficiently large to avoid numerical artifacts in the bound
single particle states due to the cell boundaries (in particular, no
artifacts from the artificial cubic symmetry of the supercell are
found). In the following we discuss three different QD sizes with
diameters $d=4.5, 5.7, 7.7$ nm and heights $h=1.6, 2.3, 3.0$ nm,
respectively. The WL thickness is one lattice constant $c$.

In a first step we neglect strain induced displacements of the atoms
and concentrate on effects which can exclusively be attributed to
crystal-field splitting and spin-orbit coupling. The influence of
strain effects will be discussed separately in
Sec.~\ref{sec:StrainBandstruc}.

In contrast to semiconductor heterostructures with a zinc blende
structure, the III-V wurtzite nitrides exhibit a considerably larger
built-in electrostatic field~\cite{BeFi97}. In order to account for
this field in our model, the electrostatic potential
$\phi_p(\mathbf{r})$ is obtained from the solution of the Poisson
equation and enters as a site-diagonal contribution $V_p(\mathbf{r})
= -e\phi(\mathbf{r})$ to the TB Hamiltonian~\cite{SaAr2002}. The
polarization $\mathbf{P}$ has two contributions, the spontaneous
polarization $\mathbf{P}^{\text{spont}}$ and the piezo-electric
contribution $\mathbf{P}^{\text{piezo}}$ caused by strain inside the
system. For the latter we apply the approximation described in
Ref.~\cite{DeRDA2004} and assume
$\mathbf{P}^{\text{piezo}}\sim\mathbf{e}_z$ , which is a reasonable
approximation for the considered QD geometry~\cite{SaAr2002}.
Further details of this procedure are given in Ref.~\cite{ScSc2006}.

\subsection{Many-Body Hamiltonian, Coulomb and Dipole Matrix
Elements} \label{sec:ManyCoulDip}

Having discussed the TB Hamiltonian used for the calculation of the
bound single-particle states, we now turn our attention to the
investigation of the optical properties of the studied QD system. We
start with the following Hamiltonian $H$ that describes the dynamics
of the interacting charge carriers in the system:
\begin{eqnarray}\label{Eq:Hamiltonian}
H &=& H_0 + H_C + H_D\,.
\end{eqnarray}
This Hamiltonian consists of three parts and is given in the basis
of the QD one-particle eigenstates. The contribution $H_0$
\begin{eqnarray*}
H_0 &=& \sum_{i} \epsilon_i^e c^{\dagger}_{i} c_{i}+\sum_{i}
\epsilon_i^h h^{\dagger}_{i} h_{i}\,, \nonumber
\end{eqnarray*}
is the one-particle part, which is diagonal in the chosen basis,
\begin{eqnarray}
\nonumber H_C & = & \frac{1}{2}\sum_{ijkl} V^{ee}_{ij,kl} \,
c^{\dagger}_{i} c^{\dagger}_{j} c_{k} c_{l} +\frac{1}{2}\sum_{ijkl}
V^{hh}_{ij,kl}\,
h^{\dagger}_{i} h^{\dagger}_{j} h_{k} h_{l}\\
 & & - \sum_{ijkl} V^{he}_{ij,kl} \, h^{\dagger}_{i} c^{\dagger}_{j}
c_{k} h_{l}\,,
\end{eqnarray}
describes the Coulomb interaction of electrons (e) in the conduction
band states and holes (h) in the valence band states, and
\begin{equation}
\label{Eq:DipolHamiltonian}
 H_D = \sum_{i,j} \,\left(
e \langle i | \mathbf{E}\mathbf{r} | j \rangle \,
c_{i}^{\phantom{\dagger}} h_{j}^{\phantom{\dagger}}\,+\,\text{h.c.}
\right)\,,
\end{equation}
includes the coupling of the electronic system to an external
electromagnetic field $\mathbf{E}$ in dipole approximation. The
creation and annihilation operators for electrons (holes) in the
single-particle state $|i\rangle$ with energy $\epsilon^{e}_i$
($\epsilon^{h}_i$) are denoted by $c^{\dagger}_i$ ($h^{\dagger}_i$)
and $c^{}_i$ ($h^{}_i$), respectively. The Coulomb interaction
matrix elements are labeled by $V^{\lambda\lambda'}_{ijkl}$.

The calculation of the Coulomb interaction matrix elements requires
-- at least in principle -- the knowledge of the localized basis
states implicitly underlying the TB wave functions. However, since
the Coulomb matrix elements are dominated by the long-range
character of the interaction, in the calculation of these matrix
elements the charge densities in the localized orbitals are
approximated by point charges. A more detailed discussion of this
issue is given in Ref.~\cite{ScSc2006}. This approximation leads to
the following explicit form of the Coulomb matrix elements:
\begin{eqnarray}\label{EqCoulappr}
&V_{ijkl}=
\sum_{\mathbf{R}\mathbf{R}'}\sum_{\alpha\beta}c_{\mathbf{R}\alpha}^{i\ast}
c_{\mathbf{R}'\beta}^{j\ast}c^k_{\mathbf{R}'\beta}c^l_{\mathbf{R}\alpha}V(\mathbf{R}-\mathbf{R}')\,,
\label{Eq:CoulmbMatrix}
\end{eqnarray}
with
\begin{eqnarray*}
&V(\mathbf{R}-\mathbf{R}')=
\frac{e^2_0}{4\pi\varepsilon_0\varepsilon_r|\mathbf{R}-\mathbf{R}'|}\quad\text{for}
\quad \mathbf{R}\not=\mathbf{R}'\label{EqCoulappr1}
\end{eqnarray*}
and
\begin{eqnarray*}
&V(0)=\frac{1}{V^2_{uc}}\int_{uc}d^3rd^3r'\frac{e^2_0}{4\pi\varepsilon_0\varepsilon_r|\mathbf{r}-\mathbf{r}'|}\approx
V_0 \,.
\end{eqnarray*}
The expansion coefficients $c^{i}_{\alpha,\mathbf{R}}$ are related
to the $\text{i}^{\text{th}}$ one-particle wave function
$\Phi_i(\mathbf{r})=\sum_{\alpha,\mathbf{R}}c^{i}_{\alpha,\mathbf{R}}\phi_{\alpha,\mathbf{R}}(\mathbf{r})$
where  $\phi_{\alpha,\mathbf{R}}(\mathbf{r})$ denotes the atomic
wave functions localized at the lattice site $\mathbf{R}$.

To calculate the dipole matrix elements
$\mathbf{d}^{eh}_{ij}\propto\langle\psi^{e}_i|\mathbf{r}|\psi^{h}_j\rangle$
entering Eq.~(\ref{Eq:DipolHamiltonian}), we use numerically
orthogonalized Slater orbitals~\cite{ScSc2006} to account for the
short-range character of the dipole operator. The numerically
orthogonalized Slater orbitals fulfill the basic properties of the
localized basis states underlying the TB model: symmetry, spatial
orientation~\cite{SlKo54}, and orthogonality. We also include the
anion-cation structure of the crystal and the slight nonlocality of
the dipole operator by including contributions from up to second
nearest neighbors.

In analogy to the bulk systems, a separation of the orbital and spin
part (both included in the index $\alpha$)  is prohibited by the
spin-orbit coupling. In contrast to the bulk case, as will be
discussed in the following section, the strong band mixing prevents
a strict classification of QD single-particle states according to
their angular momentum. Therefore, even total angular momentum
selection rules are no longer applicable. Ignoring the band mixing
characteristics in the zero-dimensional structures, any treatment of
many-body effects based on strict selection rules for the total
angular momentum, yields inaccurate predictions of level
degeneracies. However, as discussed in detail in
Ref.~\cite{BaSc2007}, the selection rules can always be analyzed on
symmetry grounds.

With the dipole and Coulomb matrix elements and the many-body
Hamiltonian, given in Eq.~(\ref{Eq:Hamiltonian}), the calculation of
optical spectra can be carried out as described in
Refs.~\cite{BaGa2004,BaSc2007}.

\section{Results for lens-shaped $\mathrm{\bf I{n}N}$ Quantum Dots}
\label{chap:resultstruncGaN}

\begin{figure}[t]
\centering
\includegraphics{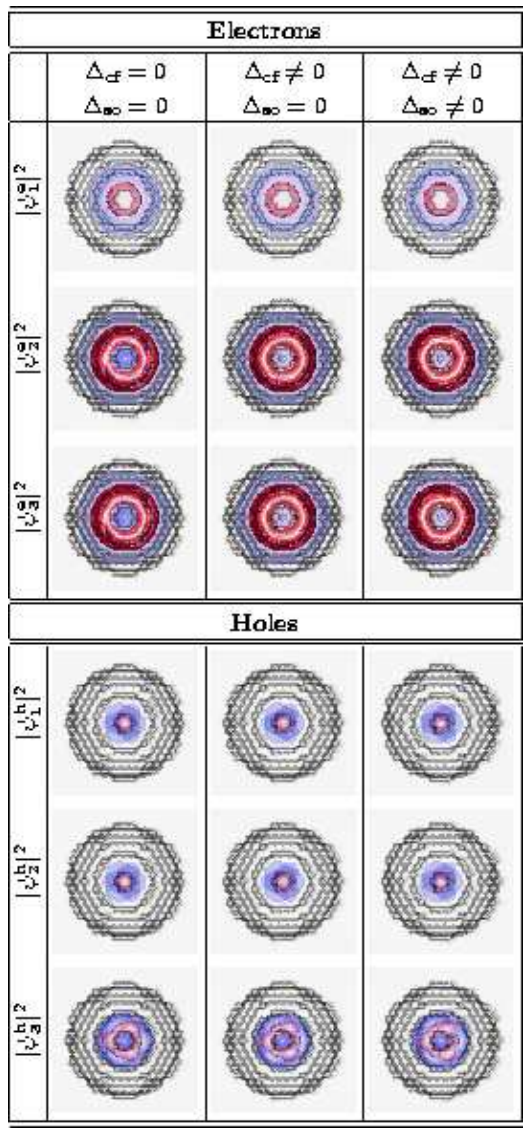}
\caption{(Color online) Top view of the large lens-shaped InN QD
structure with the first three bound states for electrons (upper
part) and holes (lower part). Depicted are isosurfaces of the
probability density with 10\% (blue) and 60\% (red) of the maximum
value.} \label{fig:oneparticle}
\end{figure}
\subsection{Single Particle states and Energies}
Having determined the TB parameters, the single-particle states and
energies of the three different QDs discussed in
Sec.~\ref{sec:geometryInN} can be calculated.

First we turn our attention to the single particle states of the
large QD ($d=7.7$ nm, $h=3.0$ nm). In order to assess the impact of
the CF splitting and SO coupling, we have performed our calculations
in three steps. In a first step we neglect both SO interaction and
CF effects ($\Delta_{\text{so}}=0, \Delta_{\text{cf}}=0$). In this
case we are left with the $sp^3$ TB model discussed in
Ref.~\cite{ScSc2006}. In step two we introduce only the crystal
field splitting ($\Delta_{\text{so}}=0, \Delta_{\text{cf}}\neq0$),
by including the additional parameter $E^A_{p_z,p_z}$ in the TB
model. In the final step both contributions are taken into account
($\Delta_{\text{so}}\neq 0, \Delta_{\text{cf}}\neq0$). In each step
the TB parameters are re-optimized in such a way that band gap and
the energetic positions of other bands of the wurtzite bulk band
structure at the $\Gamma$ point are reproduced. In other words, in
the first step ($\Delta_{\text{so}}=\Delta_{\text{cf}}=0$) we use
the parameters given in Ref.~\cite{ScSc2006}. In step two
($\Delta_{\text{so}}=0,\Delta_{\text{cf}}\neq 0$) the parameter
$E^A_{p_z,p_z}$ is included in the TB-model to reproduce the
splitting of the $A-C$ bands. Of course all the other matrix
elements are also re-adjusted to obtain the correct band structure.
Starting from these TB parameters, in the final step
($\Delta_{\text{so}}\neq 0,\Delta_{\text{cf}}\neq 0$) the additional
parameter $\lambda$ is taken into account to reproduce the splitting
of the bands $A-B$. Again, all parameters are re-adjusted to
reproduce the correct energetic positions of the different bands.
The resulting parameters are given in
Tab.~\ref{TBparametersInNwurtz}. Figure~\ref{fig:oneparticle} shows
the QD geometry and first three bound one-particle states for
electrons and holes, respectively, including the influence of the
built-in field.

According to the nodal structure, the depicted electron ground state
$\psi^{e}_1$ can be classified as $s$-like. The first two excited
states $\psi^{e}_{2}$ and $\psi^{e}_{3}$ can be classified as $p_+$
and $p_-$ states, respectively. Such a classification is not
possible for the hole states, since these states undergo strong band
mixing effects. Considering only a single valence band for the
description of the bound hole states in an InN QD is not valid. The
observation of strong band mixing effects is in agreement with other
multi-band approaches.~\cite{AnORe2000,FoBa2004,WiSc2006} From
Fig.~\ref{fig:oneparticle} we can deduce that CF splitting and SO
coupling do not alter the single-particle level structure. In other
words, for the ordering of the first three bound electron and hole
states in a lens-shaped InN QD, the contributions from SO and CF
splitting are negligible.

\begin{table}[b]
\centering \caption[Character table for the single group
$C_{3v}$.]{\label{tab:singleC3v} Character table for the single
group $C_{3v}$ (Ref.~\cite{Corn69}).}
\begin{tabular}{|c|ccc|}
\hline\hline
 & $\{E\}$ & $\{2C_3\}$ & $\{3\sigma_v\}$ \\ \hline
$\Gamma_1$ & 1 & 1 & 1 \\
$\Gamma_2$ & 1 & 1 & -1 \\
$\Gamma_3$ & 2 & -1 & 0 \\\hline\hline
\end{tabular}
\end{table}

After this discussion of the single-particle states we focus on the
single-particle energies of the large QD. In
Tab.~\ref{tab:EnergiesCFSO} the energies of the first three bound
electron and hole states under the influence of the built-in field
are displayed. From this table we conclude that the electron states
are only slightly shifted to lower energies by CF splitting and SO
coupling. Without SO coupling ($\Delta_{\text{so}}=0$), and taking
only CF splitting into account ($\Delta_{\text{cf}}\neq0$), the hole
states are shifted to higher energies. With SO coupling, the hole
energy spectrum is shifted to lower energies, compared to the case
neglecting both contributions ($\Delta_{\text{cf}}=0$,
$\Delta_{\text{so}}=0$). Additionally it turns out that the
degeneracy of the hole states $\psi^{h}_2$ and $\psi^{h}_3$
($p$-shell) is lifted when SO coupling is considered. Of course,
each state is still twofold degenerate due to time reversal
symmetry~\cite{BaPa75}. Because of the small SO energies of the bulk
materials, the splitting is rather small
($\Delta_{\psi^{h}_1,\psi^{h}_2}=1.2$ meV). The same is true for the
electron $p$-states. Here the influence of the SO coupling is even
weaker ($\Delta_{\psi^{h}_1,\psi^{h}_2}=0.1$ meV).

As discussed recently,~\cite{BaSc2007}, for the system under
investigation, a TB model which neglects the weak crystal-field
splitting and spin-orbit coupling, must result in degenerate
$p$-shells for electrons and holes. The origin of these degeneracies
is the $C_{3v}$ symmetry of the combined system of QD geometry and
underlying wurtzite lattice.

From the splitting of the hole states $\psi^{h}_2$ and $\psi^{h}_3$,
one can deduce that the spin-orbit interaction alters the symmetry
of the system. This can be understood by an analysis of the
corresponding character tables for $C_{3v}$ single and double
groups.

\begin{table}
\centering \caption[Single-particle energies for the large InN QD in
the presence and absence of crystal field splitting and spin-orbit
coupling.]{\label{tab:EnergiesCFSO}Single-particle energies for the
large InN QD in the presence and absence of crystal field splitting
and spin-orbit coupling. Each of the given states is two-fold
degenerate due to time reversal symmetry.}
\begin{tabular}{|c|c|c|c|}
\hline\hline
 & $\Delta_{\text{so}}= 0, \Delta_{\text{cf}}=0$&
$\Delta_{\text{so}}= 0, \Delta_{\text{cf}}\neq0$ &
$\Delta_{\text{so}}\neq 0, \Delta_{\text{cf}}\neq0$ \\ \hline
$E^{e}_1$ [eV] & 1.4770 & 1.4585 & 1.4557 \\
$E^{e}_2$ [eV] & 1.6660 & 1.6464 & 1.6417 \\
$E^{e}_3$ [eV] & 1.6660 & 1.6464 & 1.6418 \\\hline
$E^{h}_1$ [eV] & 0.9021 & 0.9041 & 0.8993 \\
$E^{h}_2$ [eV] & 0.9021 & 0.9041 & 0.8981 \\
$E^{h}_3$ [eV] & 0.8964 & 0.8989 & 0.8924 \\\hline\hline
\end{tabular}
\end{table}

\begin{table}[b]
\centering \caption[Character table for the double group
$\bar{C}_{3v}$.]{\label{tab:doubleC3v} Character table for the
double group $\bar{C}_{3v}$ (Ref.~\cite{KaRo73}).}
\begin{tabular}{|c|cccccc|}
\hline\hline
 & $\{E\}$ & $\{\bar{E}\}$ & $\{2C_3\}$ & $\{2\bar{C}_3\}$ & $\{3\sigma_v\}$ & $\{3\bar{\sigma}_v\}$
 \\ \hline
$\Gamma_1$ & 1 & 1 & 1 & 1 & 1 & 1 \\
$\Gamma_2$ & 1 & 1 & 1 & 1 & -1 & -1 \\
$\Gamma_3$ & 2 & 2 & -1 & -1 & 0 & 0 \\
$\Gamma_4$ & 2 & -2 & 1 & -1 & 0 & 0 \\
$\Gamma_5$ & 1 & -1 & -1 & 1 & $i$ & $-i$ \\
$\Gamma_6$ & 1 & -1 & -1 & 1 & $-i$ & $i$ \\ \hline\hline
\end{tabular}
\end{table}

\emph{Without} spin-orbit coupling, the symmetry of the system is
determined by the \emph{single} group $C_{3v}$. Looking at the
character table, shown in Tab.~\ref{tab:singleC3v}, we conclude that
this group allows for double degenerate levels, since the group
contains a two-dimensional representation $\Gamma_3$. One example of
such degenerate states are the $p$-shell states for electrons
($\psi^{e}_2$; $\psi^{e}_3$) and holes ($\psi^{h}_1$; $\psi^{h}_2$),
shown in Fig.~\ref{fig:oneparticle}. {Of course, in the absence of
the SO coupling, in addition, each state is two-fold spin
degenerate.}

Including SO coupling, one has to deal with the \emph{double} group
$\bar{C}_{3v}$. The character table of the double group is given in
Tab.~\ref{tab:doubleC3v}. This group allows only two dimensional
representations, even if the time reversal symmetry is
included~\cite{BaPa75}. More specifically, the degeneracy of the
irreducible representations $\Gamma_3$ and $\Gamma_4$ is \emph{not}
doubled by the time reversal symmetry.~\cite{BaPa75} In other words,
no four-fold degenerate state in the energy spectrum can exist.
Consequently, the electron $p$ states ($\psi^{e}_2$ and
$\psi^{e}_3$) are also no longer degenerate, but in this particular
case the splitting is only of the order of some $\mu$eV. We note
that this is in contrast to the findings in Ref.~\cite{WiSc2006},
where (at least numerically) exact degeneracy of the electronic
$p$-states has been reported.

\begin{figure}[t]
\centering
{\includegraphics{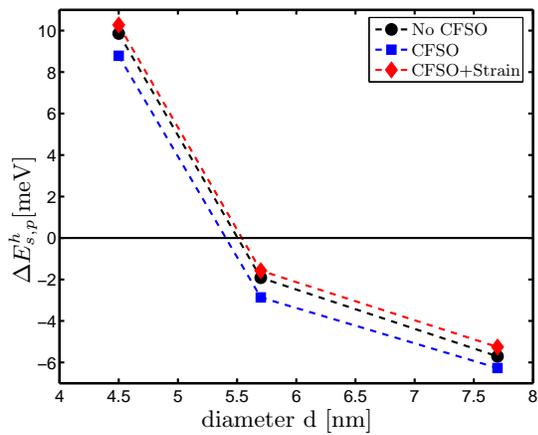}} \caption{(Color online)
The energy splitting $\Delta E^{{h}}_{s,p}=E^{{h}}_{s}-E^{{h}}_{p}$
between the $s$ and the $p$ shell for the holes is shown in the
absence (No CFSO) and presence (CFSO) of crystal field
($\Delta_{\text{cf}}$) and spin-orbit ($\Delta_{\text{so}}$)
splitting. Results are also shown, where in addition to crystal
field and spin-orbit splitting a strain field is included
(CFSO+Strain). The influence of the strain field on the electronic
properties is discussed in detail in Sec.~\ref{sec:StrainBandstruc}.
In all three cases, $\Delta E^{{h}}_{s,p}$ changes sign with
increasing QD diameter $d$, as a level reordering occurs (the dashed
lines are included as a guide to the eye). The built-in field is
included in all the calculations.}\label{fig:comparEnergysplit}
\end{figure}

A central result of the previous work~\cite{ScSc2006,BaSc2007} was
that the strong internal electrostatic field can reverse the
energetic ordering of the first three bound hole states. We find
that, for the intermediate and the largest InN QD, in the presence
of the built-in field, the ground state is formed by the twofold
degenerate $p$ states $\psi^h_1$ and $\psi^h_2$, shown in
Fig.~\ref{fig:oneparticle}. This behavior is interchanged with
decreasing QD size, where, for the smallest QD, the $s$ state
$\psi^h_3$ becomes the hole ground state. To concentrate on this
reordering of the hole $s$ and $p$ shell, the energy splitting
$\Delta E^h_{s,p}=E^h_s-E^h_p$ is displayed in
Fig.~\ref{fig:comparEnergysplit}. In order to analyze the impact of
CF splitting and SO coupling on the ordering of the hole level
structure, we calculate the splitting $\Delta E^h_{s,p}$ when both
contributions are introduced in the TB approach. Since the hole $p$
shell is no longer degenerate, we average over the single-particle
energies of the states $\psi^h_1$ and $\psi^h_2$. The energy
splitting $\Delta E^h_{s,p}$ with SO coupling and CF splitting is
also shown in Fig.~\ref{fig:comparEnergysplit}. From the comparison
with $\Delta E^h_{s,p}$ in the absence of these contributions, we
find that the SO coupling and the CF splitting have only a
negligible effect on the energy splitting $\Delta E^h_{s,p}$.
Furthermore, the ordering of the first three bound hole states is
unaffected by SO coupling and CF splitting.

In summary, the CF splitting alone cannot alter the symmetry of the
system and leads only to a small energy shift of the first three
bound electron and hole states. The SO interaction, and only this
contribution, can modify the symmetry, and lifts certain
degeneracies. However, the splitting of the electron and hole $p$
shell due to the SO coupling, is very small compared to the level
spacing of the different shells. Moreover, in the presence of SO
coupling and CF splitting, one obtains the same level ordering of
the energetically lowest hole states ($s$- and $p$-shell) as in the
case were these contributions are not taken into account. This
analysis indicates that it is well justified to neglect these small
corrections of the CF splitting and the SO coupling in the system
under consideration.

\subsection{Influence of strain}
\label{sec:StrainBandstruc}

So far we have neglected the influence of the lattice mismatch
between InN and GaN. The lattice mismatch leads to the appearance of
a strain field in the nano\-structure. This field modifies the
energies of the bound electron and hole states.

For the electron states, caused by the underlying wurtzite lattice
and the assumed QD geometry, the strain field produces only an
energy shift of the bound single particle states~\cite{FoBa2003}. In
contrast to QDs with a zinc blende structure~\cite{BeZu2005}, no
degeneracies are lifted by the strain field. The situation is more
complicated for the hole states. The strain field modifies the local
valence band edges, and can therefore lead to a splitting of the
different energy bands. As discussed in Ref.~\cite{WiSc2006}, due to
the biaxial strain in the basal plane, the first two valence bands
($A$ and $B$) are shifted to higher energies, compared to the
unstrained material, whereas the third valence band ($C$ band) is
shifted to lower energies. These energy shifts may also increase the
influence of spin-orbit coupling and crystal field splitting on the
bound single particle states.

\begin{table}
\caption{Material parameters for GaN and InN. If not indicated
otherwise, all parameters are taken from Ref.~\cite{ShGa2003}.}
\begin{tabular}{ccccc}
\hline\hline Parameter & & GaN & & InN\\ \hline
a (\AA) & & 3.189 & & 3.545 \\
c (\AA) & & 5.185 & & 5.703 \\
$\Delta_{\text{so}}$ (eV) & & 0.010 (Ref.~\cite{VuMe2003}) & & 0.005 (Ref.~\cite{VuMe2003})\\
$\Delta_{\text{cf}}$ (eV) & & 0.017 (Ref.~\cite{VuMe2003}) & & 0.040 (Ref.~\cite{VuMe2003})\\
$C_{13}$ (GPa) & & 11.4 & & 9.4 \\
$C_{33}$ (GPa) & & 38.1 & & 20.0 \\
$(a_c-D_1)$ (eV) & & -9.6 (Ref.~\cite{PeMc2005}) & & -9.6 (Ref.~\cite{PeMc2005}) \\
$(a_c-D_2)$ (eV) & & -8.2 (Ref.~\cite{PeMc2005}) & & -8.2 (Ref.~\cite{PeMc2005}) \\
$D_3$ (eV) & & 1.9 (Ref.~\cite{PeMc2005}) & & 1.9 (Ref.~\cite{PeMc2005})\\
$D_4$ (eV) & & -1.0 (Ref.~\cite{PeMc2005}) & & -1.0 (Ref.~\cite{PeMc2005})\\
$a_c$ (eV) & & -4.9 (Ref.~\cite{VuMe2003}) & & -3.5
(Ref.~\cite{VuMe2003})\\\hline\hline
\end{tabular}
\label{tab:materialpara}
\end{table}

To investigate the influence of these shifts and the effect of the
possible valence band splittings, we proceed in the following way:

Since the TB parameters are fitted to the bulk band structure, we
re-calculate the bulk band structure of InN in the presence of a
strain field. To obtain the strain dependent valence band edge we
apply~\cite{ChCh96}
\begin{eqnarray}
E_1 &=&
\Delta_{\text{cf}}+\frac{1}{3}\Delta_{\text{so}}+\theta_\epsilon+\lambda_\epsilon\,,\\
E_{2,3} &=&
\frac{\Delta_\text{cf}}{2}-\frac{\Delta_\text{so}}{3}+\frac{\theta_{\epsilon}}{2}+\lambda_\epsilon\\
& &
\mp\sqrt{\left(\frac{\Delta_\text{cf}-\frac{\Delta_\text{so}}{3}+\theta_{\epsilon}}{2}\right)^2+\frac{2}{9}\left(\Delta_{\text{so}}\right)^2}\,,
\end{eqnarray}
where $\theta_\epsilon$ and $\lambda_\epsilon$ are given by
\begin{eqnarray*}
\theta_\epsilon &=&
D_3\epsilon_{zz}+D_4\left(\epsilon_{xx}+\epsilon_{yy}\right)\,\,
,\\
\lambda_\epsilon &=&
D_1\epsilon_{zz}+D_2\left(\epsilon_{xx}+\epsilon_{yy}\right)\,\, .
\end{eqnarray*}
Here, $D_1$, $D_2$, $D_3$ and $D_4$ are the valence band deformation
potentials and $\epsilon_{ii}$ denotes the diagonal components of
the strain tensor $\underline{\underline{\epsilon}}$. For the
components $\epsilon_{ii}$ we assume:
\begin{equation}
\epsilon_{zz}=\frac{a_0-a}{a}\,\, , \,\,
\epsilon_{xx}=\epsilon_{yy}=-\frac{2C_{13}}{C_{33}}\epsilon_{xx}\,\,
,
\end{equation}
where $a_0$ and $a$ are the lattice constants of the substrate and
the QD material, respectively, and $C_{13}$ and $C_{33}$ are the
stiffness constants. Strictly speaking, these equations apply only
to a quantum well, since they neglect shear strain components and
the fact that the strain field in a QD is position dependent.
However, due to the symmetry of the QD structure and the underlying
wurtzite lattice, the strain field can not lift degeneracies. This
is also confirmed by results of Winkelnkemper \emph{et.
al}~\cite{WiSc2006} for an $\text{In}_{x}\text{Ga}_{1-x}\text{N}$ QD
with a comparable symmetry. Furthermore, the hole states, for which
the modification of the valence band edge is of major importance,
are strongly localized in the region of the QD. Therefore, the
variation of the strain field outside the QD is of minor importance.
In the region of the InN QD, we use the lattice constant $a$ of GaN,
to take into account, that the QD is pseudomorphically grown on the
GaN substrate.

\begin{figure}[t!]
\centering
\includegraphics{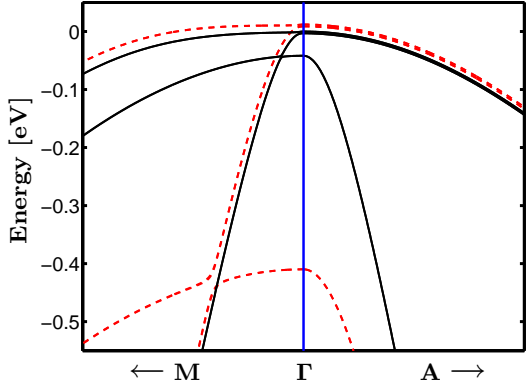}
\caption{(Color online) Calculated TB valence band structure in the
vicinity of the $\Gamma$ point. Unstrained (black solid lines) and
under compressive biaxial strain in the basal plane with elastic
relaxation along the $[0001]$ direction (red dashed lines). The
splitting between the $A$ and $B$ valence bands at the $\Gamma$
point is $\approx 3.1$ meV ($\approx 3.1$ meV) in unstrained
(strained) InN.} \label{fig:bulkstrainInN}
\end{figure}

According to Ref.~\cite{ChCh96}, the band gap shift is given by
\begin{equation}
\Delta
E^{\text{gap}}=E^{\text{gap}}_0+P_{c\epsilon}-\left(\theta_\epsilon+\lambda_\epsilon\right)\,,
\end{equation}
with
\begin{eqnarray*}
P_{c\epsilon}=a_{c}\epsilon_{zz}+a_c\left(\epsilon_{xx}+\epsilon_{yy}\right)\,,\\
\lambda_{\epsilon}=D_1\epsilon_{zz}+D_2\left(\epsilon_{xx}+\epsilon_{yy}\right)\,,
\end{eqnarray*}
where $a_c$ denotes the conduction band deformation potential. The
different parameters are listed in Tab.~\ref{tab:materialpara}. The
TB parameters are readjusted to reproduce the bulk band structure in
the vicinity of the $\Gamma$ point. The resulting valence band
structure in the vicinity of the $\Gamma$ point is depicted in
Fig.~\ref{fig:bulkstrainInN}. The band structure exactly reflects
the behavior which is expected for the local band structure in the
region of the QD:~\cite{WiSc2006} the first two valence bands ($A$
and $B$) are shifted to higher energies while the third band ($C$)
is shifted to lower energies. The calculated TB parameters are now
used to investigate the influence of strain effects on the
electronic states.

Starting from these new TB parameters one can  re-calculate the
single-particle states and energies of the three different QDs
discussed in the preceding section, taking into account both CF and
SO splitting. First we focus on the single-particle energies of the
large QD ($d=7.7$ nm, $h=3.0$ nm). In Tab.~\ref{tab:EnergiesStrain},
the energies of the first three bound electron and hole states,
including the built-in field are displayed. Each of the given states
is two-fold degenerate due to time reversal symmetry. The results in
the absence and in the presence of the strain effects are compared.
First of all, the strain merely produces an energy shift of the
single particle states. {The strain field shifts both electron
states and hole states to higher energies.} This behavior reflects
the local band edge shifts of conduction and valence bands.
Following the discussion of the previous section, no four-fold
degenerate states can exist {taking SO coupling into account}. As
already discussed, the first two excited states $E^{e}_2$ and
$E^{e}_3$ are nearly degenerate in the absence of the strain field.
Also in the presence of the strain field, these states are split by
the SO coupling by $0.1\,\mathrm{meV}$ only. This analysis shows
that the splitting of the electron $p$-states is not altered by the
strain field and remains very small compared to the energy
separation of electron $s$- ($\psi^{e}_{1}$) and $p$-shell
($\psi^{e}_{2}$ and $\psi^{e}_{3}$). The splitting of the hole
states $\psi^{h}_{1}$ and $\psi^{h}_{2}$ is nearly unaffected by the
strain field.

\begin{table}[t]
\centering \caption[Single-particle energies for the large InN QD in
the presence and absence of strain
effects.]{\label{tab:EnergiesStrain} Single-particle energies for
the large InN QD in the presence and absence of strain effects. Each
of the given states is two-fold degenerate due to time reversal
symmetry. The internal electrostatic field is included in the
calculation.}
\begin{tabular}{|c|cc|}
\hline\hline
 & \multicolumn{2}{c|}{$\Delta_{\text{so}}\neq 0, \Delta_{\text{cf}}\neq
 0$}\\ \hline
 & Without Strain &
With Strain  \\
\hline
$E^{e}_1$ [eV] & 1.4557 & 1.7740 \\
$E^{e}_2$ [eV] & 1.6417 & 1.9625 \\
$E^{e}_3$ [eV] & 1.6418 & 1.9626 \\\hline
$E^{h}_1$ [eV] & 0.8993 & 0.9126 \\
$E^{h}_2$ [eV] & 0.8981 & 0.9118 \\
$E^{h}_3$ [eV] & 0.8924 & 0.9069 \\\hline\hline
\end{tabular}
\end{table}

Additionally, the level ordering is \emph{not} modified by the
strain field. In the presence of the strain field, the intrinsic
electrostatic field still reverses the energetic ordering of the
first three bound hole states. The energy splitting
$\Delta^{h}_{s,p}$ between the hole $s$ and $p$ shell including the
strain field is displayed in Fig.~\ref{fig:comparEnergysplit}. We
compare these results with the other results shown in
Fig.~\ref{fig:comparEnergysplit}, where the strain field is absent.
In conclusion, the strain field is of minor importance for the
energy splittings, and, most importantly, the ordering of the first
three bound hole states is unaffected.

\subsection{Excitonic absorption spectra}

\begin{figure}[t]
\centering
\includegraphics[totalheight=7.5cm]{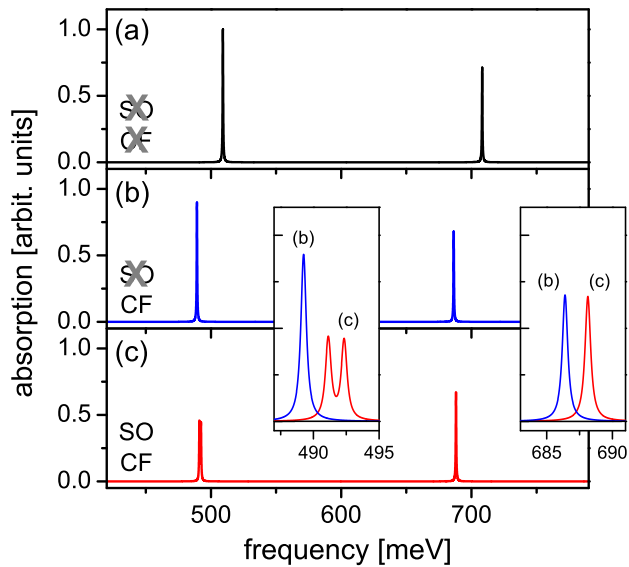}
\caption{\label{fig:ComparisonXabsorption} (Color online) Excitonic
absorption spectra for the largest InN QD in the absence of crystal
field and spin-orbit splitting (a), in the presence of crystal field
splitting and absence of spin-orbit coupling (b) and in the presence
of crystal-field and spin-orbit splitting (c). The insets show the
same data for frequencies close to the absorption peaks in (b) and
(c).}
\end{figure}

In Sec.~\ref{sec:ManyCoulDip} we have discussed the calculation of
dipole and Coulomb matrix elements. The evaluation of excitonic
absorption spectra in this section can be performed starting from
the many-particle Hamiltonian, Eq.~(\ref{Eq:Hamiltonian}), in second
quantization as given in Sec.~\ref{sec:ManyCoulDip}. For the
localized states configuration-interaction calculations are
performed. For the sake of simplicity, only the first three bound
electron and hole states are included. This can be justified by
their energy separation to higher shells in the structure. The
excitonic absorption spectra are calculated using Fermi's golden
rule.~\cite{BaGa2004}

Figure~\ref{fig:ComparisonXabsorption}~(a) shows the excitonic
absorption spectrum for the largest InN QD in the absence of CF and
SO splitting, while Fig.~\ref{fig:ComparisonXabsorption}~(b)
displays the spectrum in presence of the CF splitting but in the
absence of SO coupling. In Fig.~\ref{fig:ComparisonXabsorption}~(c),
the absorption spectrum in the presence of both CF and SO splitting
is depicted. The different absorption lines in each spectrum
correspond to the excitation of an exciton in the QD. In case (a)
and (b) the peak on the low energy side corresponds to transitions
where the electron is mainly in the ground state $\psi^{e}_1$, and
the hole is mainly in the states $\psi^{h}_1$ and $\psi^{h}_2$.
Since the hole states $\psi^{h}_1$ and $\psi^{h}_2$ are degenerate
one obtains only a single peak on the low energy side. Due to the SO
coupling, the states $\psi^{h}_1$ and $\psi^{h}_2$ are split by
about $1.2$ meV. This splitting in the single particle states
results in two lines on the low energy side in
Fig.~\ref{fig:ComparisonXabsorption}~(c) (see also the inset). Since
the splitting in the single-particle states $\psi^{h}_1$ and
$\psi^{h}_2$ is very small, the splitting of the two peaks on the
low energy side is also very small. This emphasizes again, that in
the system under consideration the SO coupling and the CF splitting
introduce only negligible corrections to the excitonic spectrum. Due
to the symmetry of the QD and the underlying wurtzite structure,
there is no polarization anisotropy. This is in contrast to
lens-shaped InAs QDs with a zinc blende structure, as discussed in
Ref.~\cite{WiWa2000}, where the $C_{2v}$ symmetry leads to
polarization anisotropy.

The peak on the high energy side mainly corresponds to the
excitation of the hole in the state $\psi^{h}_3$ and the states
$\psi^{e}_2$ and $\psi^{e}_3$. Since the electron states
$\psi^{e}_2$ and $\psi^{e}_3$ are exactly degenerate in cases (a)
and (b) and nearly degenerate in case (c), only a single line is
visible on the high energy side (cf. inset of
Fig.~\ref{fig:ComparisonXabsorption}).

Including strain effects, we obtain nearly the same splitting of the
states $\psi^{h}_1$ and $\psi^{h}_2$. Furthermore, the electron
states $\psi^{e}_2$ and $\psi^{e}_3$ are still nearly degenerate.
Only the single-particle energy gap is enlarged by the strain field
and therefore the whole excitonic absorption spectrum is shifted to
higher energies. The excitonic absorption spectrum (not shown)
resembles the spectrum in Fig.~\ref{fig:ComparisonXabsorption}~(c).

\section{Conclusion}

In this work we have presented an atomistic TB calculation of the
electronic and optical properties of lens-shaped InN QDs. We focused
our attention on the influence of the crystal-field splitting and
the spin-orbit coupling on the electronic structure as well as on
the optical properties. As it turns out, only the spin-orbit
coupling lifts certain degeneracies in the single particle spectrum.
This result is confirmed by the inspection of the character table of
the double group $\bar{C}_{3v}$, which reveals that no four fold
degenerate state can exist. However, from our calculations we obtain
only small splittings in the single-particle spectrum as well as in
the excitonic absorption spectrum. Our results indicate that it is a
reasonable assumption to neglect spin-orbit coupling and crystal
field splitting in case of lens-shaped InN QDs with a wurtzite
structure.

We do not rule out that the importance of spin-orbit coupling and
crystal-field splitting may depend on the specific system under
investigation even within the group-III nitride material system
where both effects are intrinsically weak.

\begin{acknowledgements}
We thank Paul Gartner, Norman Baer, and
Frank Jahnke for various helpful discussions. This work has been
supported by the Deutsche Forschungsgemeinschaft (research group
``Physics of nitride-based, nanostructured, light-emitting
devices'', project Cz 31/14-1,2). S. Schumacher was further
supported by the Deutsche Forschungsgemeinschaft through project
No.~SCHU~1980/3-1. We also acknowledge a grant for CPU time from the
NIC at the Forschungszentrum J\"ulich.
\end{acknowledgements}

\bibliography{./phdstef}
\bibliographystyle{apsrev}

\end{document}